\documentclass[twocolumn,showpacs,preprintnumbers,amsmath,amssymb]{revtex4}

\usepackage{graphicx}
\usepackage{dcolumn}
\usepackage{bm}

\begin{document}

\preprint{APS/123-QED}

\title{Large positive magneto-resistance in high mobility 2D electron gas:\\ interplay of short and long range disorder}

\author{V. Renard$^{a,c}$, Z. D Kvon $^{b}$, G. M Gusev$^{d}$, J.C
Portal$^{a,c,e}$}
 \affiliation{$^{a}$ GHML, MPI-FKF/CNRS, BP 166, F-38042, Grenoble Cedex 9, France\\
 $^{b}$Institute of Semiconductor Physics, 630090 Novosibirsk, Russia\\
 $^{c}$INSA 135, Avenue de Rangueil 31 077 Toulouse Cedex 4, France\\
 $^{d}$Instituto de Fisica da Universidade de Sao Paulo, SP, Brazil\\
 $^{e}$Institut Universitaire de France, Toulouse France}

\date{\today}

\begin{abstract}
We have observed a large positive quasi-classical
magneto-resistance (MR) in a high mobility 2D electron gas in
AlGaAs/GaAs heterostructure. The magneto-resistance is
non-saturating and increases with magnetic field as $\rho_{xx}\sim
B^{\alpha} (\alpha=0.9-1.2)$. In antidot lattices a non-monotonic
MR is observed. We show that in both cases this MR can be
qualitatively described in terms of the theory recently advanced
by Polyakov \textit{et al} (PRB, \textbf{64}, 205306 (2001)).
Their prediction is that such behavior as we observe may be the
consequence of a concurrent existence of short and long range
scattering potentials.
\end{abstract}

\maketitle

\section{\label{sec:level1}Introduction}

In recent years, there has been a revival of interest for the
semi-classical transport properties of a two-dimensional electron
gas (2DEG) particularly for the behavior of the classical
magneto-resistance (MR). On one hand, it is due to the
technological progress in the preparation of high mobility
heterostructures, 2DEGs with artificial scattering and modulating
potential, antidot lattices, weak modulated one-dimensional
structures, and so on. On the other hand, a remarkable progress
has been made recently in the theoretical understanding of the
importance of memory effects in semi-classical magneto-transport.
In the simplest situation of a degenerated non-interacting 2DEG
with isotropic Fermi surface the conventional Boltzmann-Drude
approach yields zero magneto-resistance, i.e. the value of the
longitudinal component of the resistivity tensor $\rho_{xx}$ does
not depend on magnetic field. However, already in the pioneering
work of Ref.~\onlinecite{Baskin1} the importance of non-Boltzmann
classical memory effects on the magneto-transport of 2DEGs with a
small a number of impurities was demonstrated and a large negative
MR was predicted. More than ten years later this MR was observed
in experiments performed in antidot lattices$^{2-5}$. However it's
only recently that the purely classical origin of this MR was
fully recognized. This MR may be either negative or positive
depending on the type of the scattering potential, and it appears
as a consequence of memory effects$^{6-11}$ both in high (
$\omega_c>1$)$^{6-10}$ and weak magnetic fields
($\omega_c<1$)$^{11}$. Recently a very important step in the
understanding of the role of memory effects on the
magneto-transport of a real 2DEG in strong magnetic field was done
in Ref.~\onlinecite{Polyakov}. In this paper the theory of the
magneto-resistance of a 2DEG scattered by a short -range disorder
potential in the presence of a long-range correlated random
potential has been advanced. The most important result of this
theory is that the interplay of two types of scattering potential
generates a new behavior of $\rho_{xx}(B)$ absent when only one
kind of disorder is present, and leads to a positive
non-saturating MR.

The purpose of this paper is to compare the experimental behavior
of the quasi-classical MR in high mobility 2DEGs in AlGaAs/GaAs
heterostructure with and without artificial scatters to the
theory$^{9}$. We have observed a large positive non-saturating MR
of the 2DEG and the addition of artificial scatters (antidots) in
this 2DEG results in a negative MR in intermediate fields that
leads a non-monotonic dependence of $\rho_{xx}(B)$ predicted in
Ref.~\onlinecite{Polyakov}.

\section{Experimental set up}

The samples we studied were: 1) high mobility 2DEGs in MBE grown
AlGaAs/GaAs heterostructures with a spacer thickness of 40 nm
(sample 188 and 189) and 60 nm (sample 218); 2) lattices of
antidot fabricated on the basis of this 2DEG. The parameters of
the 2DEG at $T$= 4.2 K were the following: sample 188 -
$\mu=(5-10)*10^5 cm^2/Vs$ at electron densities $N_s=(3-5)*10^{11}
cm^{-2}$; sample 189 - mobility $\mu=1.2*10^6 cm^2/Vs$ at
$N_s=6*10^{11}cm^{-2}$ and sample 218 - $\mu=(3-6)*10^5 cm^2/Vs$
at electron densities $N_s=(1-2)*10^{11} cm^{-2}$.The samples were
photolithographically processed into 50 $\mu$m Hall bars with the
distance between the four voltage probes on each side 100, 250 and
100 $\mu$m and Ohmic contacts to the Hall bars were prepared by
annealing of AuNiGe. An antidot lattice with the period $d$=0.6
$\mu$m was fabricated on the basis of sample 218 with Hall bars on
it by means of electron lithography and consequent plasma etching.
The samples were measured at $T=(1.5-40) K$ in magnetic fields up
to 15 T using superconducting magnet and a Variable Temperature
Insert. The data was acquired using standard low frequency lock-in
technique.

\section{Results and analysis}

\begin{figure}
\includegraphics[angle=-90,width=\columnwidth]{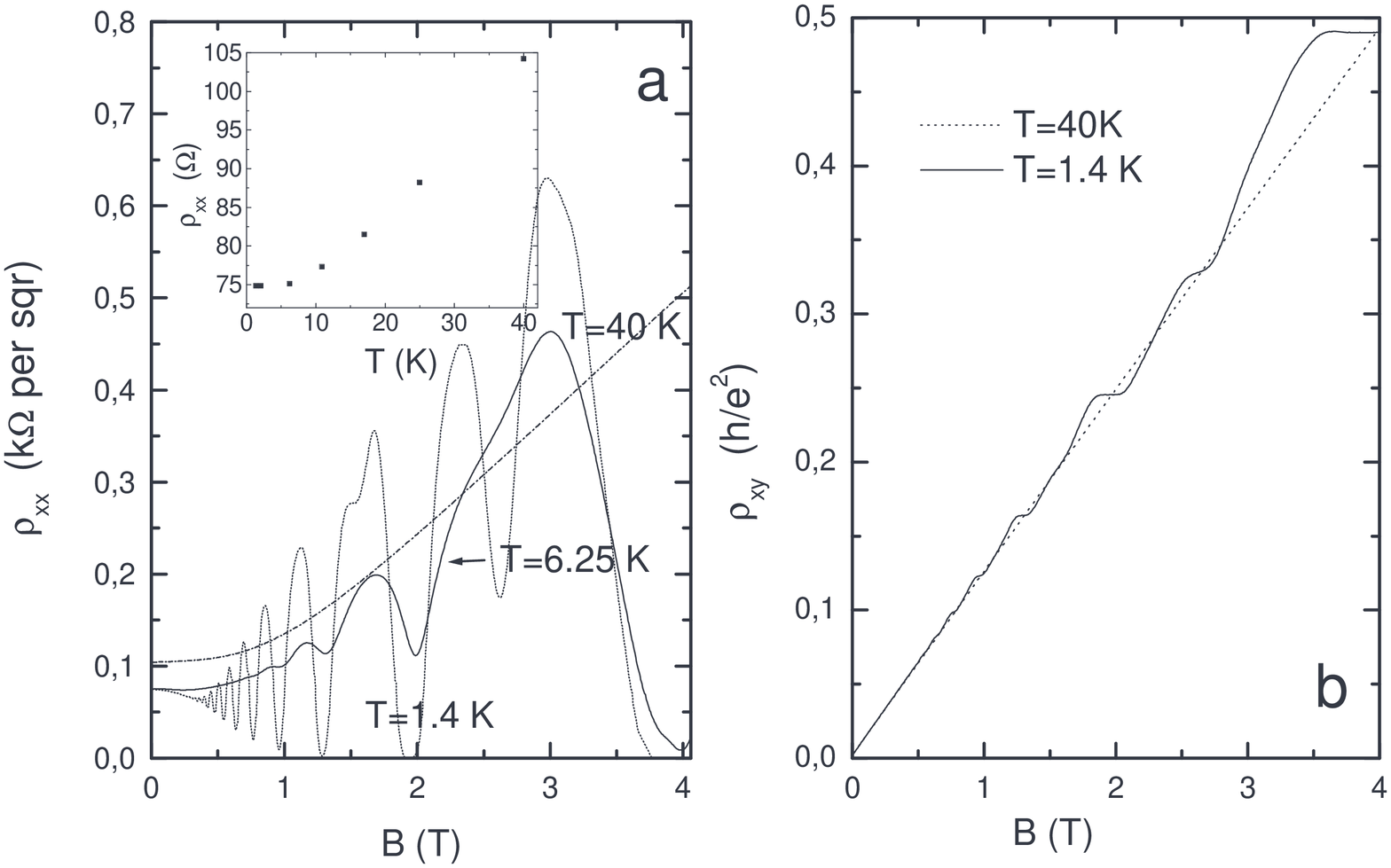}
\caption{\label{fig:fig1} Dissipative (a) and Hall resistivity (b)
of sample 218 at different temperatures. Insert to (a) show the
zero-field $T$-dependence of $\rho_{xx}$}
\end{figure}

Fig.~\ref{fig:fig1}a shows $\rho_{xx}(B)$ for the sample 218 at
different temperatures in magnetic fields up to 4 T at the
electron density $N_s = 1.9*10^{11} cm^{-2}$ (The mobility is
$\mu=4.3*10^5 cm^2/Vs$ at  4.2 K so that $\omega_c\tau= 215$ at 5
T and the mobility is $\mu=3*10^5 cm^2/Vs$ at 40 K with
$\omega_c\tau=150$ at 5 T). The Hall resistance $\rho_{xy}(B)$
under the same conditions is presented in Fig.~\ref{fig:fig1}b.

\begin{figure}
\includegraphics[angle=-90,width=\columnwidth]{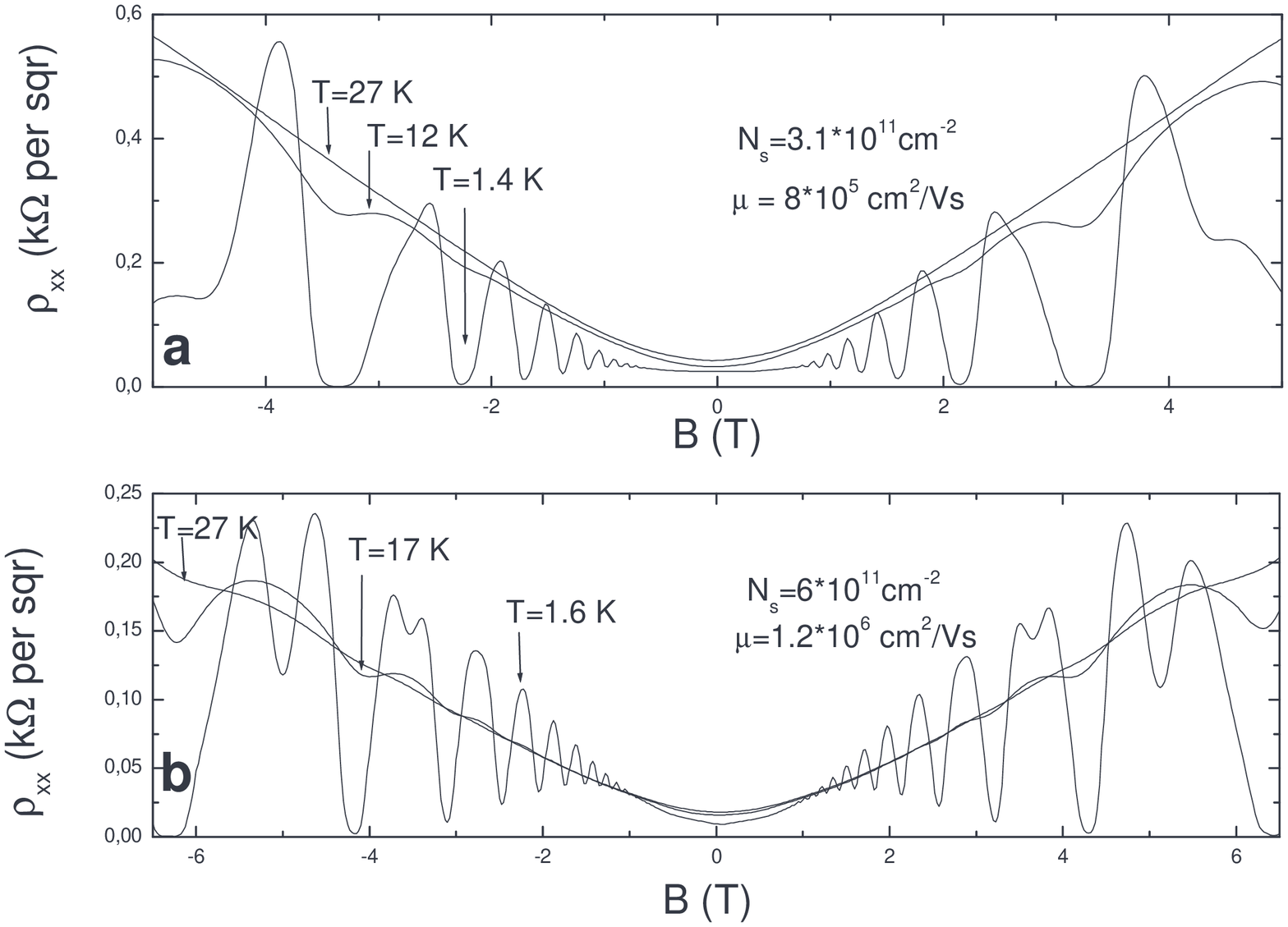}
\caption{\label{fig:fig2} Dissipative resistivity at different
temperatures for sample 188 (a) and 189 (b)}
\end{figure}

At the lowest temperature ($T=1.4$ K) the conventional behavior
for high mobility 2DEGs is observed : For $B<1$ T there is a small
parabolic-like negative MR (possibly due the interaction
effects$^{13,14}$ not discussed here), then the Subnikov de Haas
(SdH) oscillations appear, and finally the Quantum Hall Effect
sets in with deep minima in $\rho_{xx}(B)$ and corresponding
plateaus in $\rho_{xy}(B)$. For high temperatures we observe the
disappearance of the SdH oscillations and the magneto-resistance
turning to $\Delta\rho_{xx}(B)/\rho_{0}\sim B^\alpha$ with
$\alpha$ close to 1.\\The same behavior (except for the low field
range where there is no parabolic NMR) is observed in samples 188
and 189 with larger electron densities and mobilities
($N_s=3.1*10^{11} cm^{-2}$, $\mu=8*10^5 cm^2/Vs$ and
$N_s=6*10^{11} cm^{-2}$ $\mu=1.2*10^6 cm^2/Vs$ at $T=4.2$ K
respectively for 188 and 189. See Fig.~\ref{fig:fig2}a and
~\ref{fig:fig2}b). The value of $\alpha$ is close to the value
found for sample 218. \\A large non-saturating MR was actually
observed in earlier papers devoted to high mobility ($\mu>10^5
cm^2/Vs$) 2DEGs$^{15,16}$. However all measurements were performed
at temperatures $T\leq4.2$ K and the semi-classical MR was not
clearly detected because of the oscillating dependence of
$\rho_{xx}(B)$. Our aim here is to study the MR dependence in
strong magnetic field ($\omega_c\tau\gg 1$) in the high
temperatures quasi-classical regime (i.e. $\hbar\omega_c/2\pi<kT,
kT\ll E_F, N>1$, $N$ is the number of occupied Landau level).
According to the usual conception of the magneto-transport in high
magnetic field the increase of temperature should lead to: 1) a
decrease in the amplitude of SdH oscillations and their complete
suppression for $kT>\hbar\omega_c/2\pi$ and 2) a weak magnetic
field dependence of the resistance as long as $kT\ll E_F$. In this
regime $\Delta\rho_{xx}(B)/\rho_{0}\approx(kT/E_F)^{2}$ so we
should not observe a value of $\Delta\rho_{xx}(B)/\rho_{0}$
exceeding 10\% even at 5 T.\\Fig.~\ref{fig:fig1}a and
Fig.~\ref{fig:fig2} show that our experiment completely supports
the first point: Heating the samples to temperatures higher than
20 K leads to the complete disappearance of SdH oscillations.
However our high-T experiments reveals
$\Delta\rho_{xx}(B)/\rho_{0}$ as high as 10-15 for all
our samples, a clear contradiction with the second assumption.\\
Note that the value of the Fermi energy is $E_F=8 meV$ for sample
218, $E_F=16 meV$ for sample 188 and $E_F=31 meV$ for sample 189
which are significantly larger than $kT$ even at the highest
temperature 40 K ($kT=3.5 meV$ at 40 K). So that the MR can be not
attributed to a non-degeneracy of the 2DEG at this temperature.
More over, as can be clearly seen from Fig.~\ref{fig:fig2} the
curves are very symmetric and the possibility of mixing between
Hall and longitudinal signal can be excluded.

\begin{figure}
\includegraphics[angle=-90,width=\columnwidth]{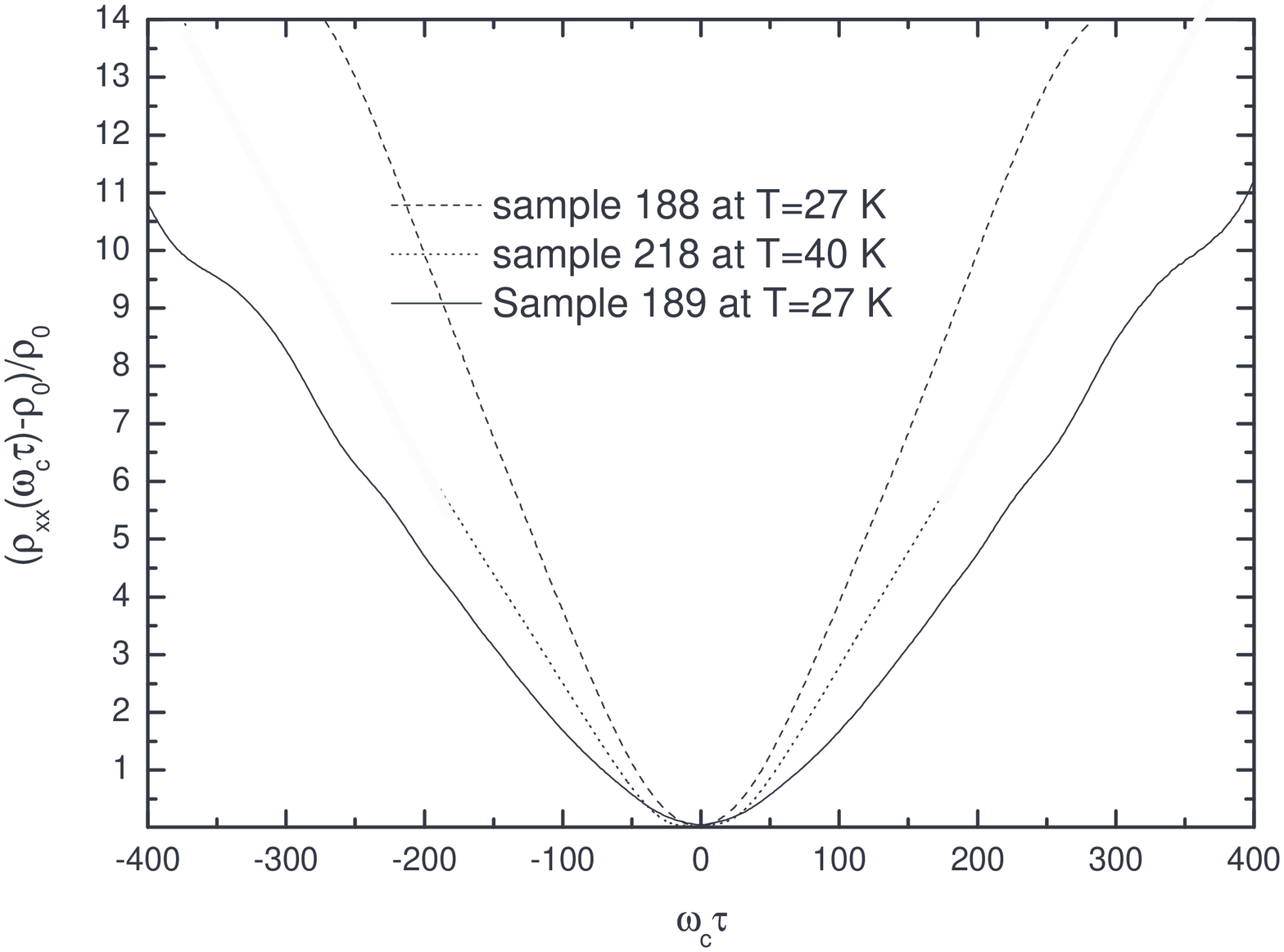}
\caption{\label{fig:fig3} Magnetoresistance
$\Delta\rho_{xx}(\omega_c)/\rho_{0}$ in the quasi-classical regime
for all three samples}
\end{figure}

Fig.~\ref{fig:fig3} summarizes the results for all samples :
$\Delta\rho_{xx}(\omega_c)/\rho_{0}$ are plotted (the curves are
displayed in order to keep the condition $N>1$ satisfied).

At first glance this result is not surprising; indeed a theory of
magneto-transport in a long range scattering potential (which is
the case for high mobility 2DEGs in AlGaAs/GaAs heterostructures)
has recently been proposed$^{8}$. It predicts a positive MR.
However according to this theory $\Delta\rho_{xx}(B)/\rho_{0}<3$
and, more importantly, $\Delta\rho_{xx}(B)/\rho_{0}$ should show a
maximum around $\omega_c\tau=(100-200)$ and go to zero in the
limit of infinite fields (this is the sign of the localization of
electrons in this limit).\\Fig.~\ref{fig:fig3} shows that we
observe no saturation even at $\omega_c\tau$ as high as
$(3-4)*10^2$. So we are in the interesting following situation:
Theoretically it is predicted that if only one kind of scattering
potential is present (short range or long range) there should be
either no MR at all or a negative MR in the limit of infinite
magnetic field. On the contrary we observe a large
($\Delta\rho_{xx}/\rho_{0}\sim 10$) non-saturating
magneto-resistance. Note that quantum calculation in the framework
of Born approximation$^{12}$ can give a positive MR at high
temperatures (when SdH oscillations are washed out) but the value
of the exponent $\alpha$ is then 0.5 not around 1 as we observe so
the theoretical framework completely fails to explain our data.
\\A qualitative explanation of our results might be given in the recent work of
Ref.~\onlinecite{Polyakov} where the theory of a semi-classical
magneto-transport in 2DEGs with both kinds of disorder potential
(which is always the case in any sample) has been advanced. The
more important conclusion of this theory is that the interplay of
short range disorder (random ensemble of impenetrable discs of
diameter $a$ and density $n$) and long-range correlated disorder
generates a new behavior of $\rho_{xx}(B)$ including the
appearance of a non-saturating positive semi-classical MR. The
autor shows that the localization ($\rho_{xx}\rightarrow 0$ when
$B\rightarrow\infty$) induced by the presence of only one kind of
disorder is destroyed by the presence the other kind of disorder.
In this theory the short range scattering is characterized by the
mean free path $l_s$ and the long range scattering by $l_L$. In
the hydrodynamic limit ($a\rightarrow 0, n\rightarrow\infty,
l_s=const$) the theory gives four ranges of magnetic field
dependence for the positive MR. In the first range
$\rho_{xx}=const$, as $B$ is increased $\rho_{xx}$ turns to
$\rho_{xx}\sim B^{12/7}$, then to $\rho_{xx}\sim B^{10/7}$ and
finally to $\rho_{xx}\sim B^{10/13}$. Qualitatively the experiment
shows a similar behavior (see Fig.~\ref{fig:fig1} and
Fig.~\ref{fig:fig2}). But a more detailed comparison shows that
there are some differences: For the sample 218 we observe a first
range with $\rho_{xx}=const$ and then a transition to the
dependence $\rho_{xx}\sim B^{\alpha}$ with $\alpha=0.9-1.1$
increasing with the magnetic field (this value is slightly larger
than 10/13). For the sample 188 and 189 the picture is different:
at weak magnetic fields there is no $\rho_{xx}=const$ range,
instead we observe $(\rho_{xx}/\rho_0-1)\sim B^{2}$ and at
magnetic fields where $\rho_{xx}(B)/\rho_0\gg1$ the behavior is
the same as for sample 218. There is nothing surprising in the
observed difference between the experiment and the theory because
the hydrodynamic limit is a very idealized picture and the
scattering potential in a real high mobility 2DEG is much more
complicated. Also, note that at temperatures $T\geq20$ K a
significant contribution (see the insert to Fig.~\ref{fig:fig1}a)
of the phonon scattering appears that can modify the magnetic
field dependence of $\rho_{xx}$. Obviously phonon scattering
should not radically change the situation principally because
phonon scattering plays the same role as short range scattering.
Nevertheless an accurate description of the experiment at high
temperatures requires a theory including phonon
scattering.\\\\
A very interesting prediction of the theory is given for antidot
arrays. In this case the theory predicts a negative MR in
intermediate magnetic fields leading to a non-monotonic dependence
of $\rho_{xx}(B)$.

\begin{figure}
\includegraphics[angle=-90,width=\columnwidth]{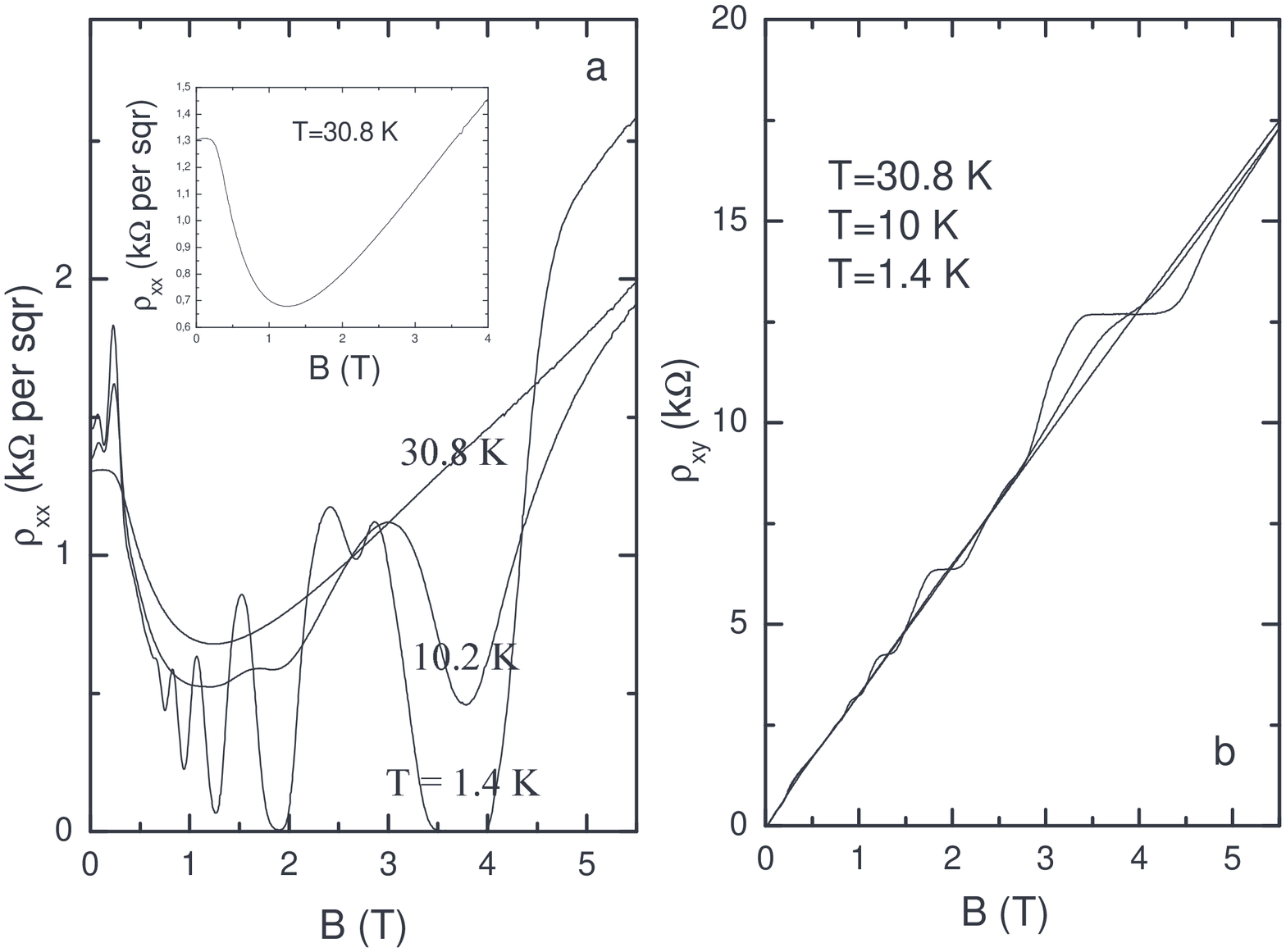}
\caption{\label{fig:fig4}Dissipative (a) and Hall resistivity (b)
of the antidot lattice at different temperatures }
\end{figure}

Fig.~\ref{fig:fig4}a shows the experimental dependence
$\rho_{xx}(B)$ measured in an antidot lattice, fabricated on the
basis of sample 218 at temperatures between 1.4 K and 40 K for
$N_s=1.8*10^{11} cm^{-2}$. The Hall resistance measured under the
same conditions is shown in Fig.~\ref{fig:fig4}b. The insert to
Fig.~\ref{fig:fig4}a reveals that at high enough temperatures
(when SdH oscillations are suppressed) the experimental curve
$\rho_{xx}(B)$ demonstrate a non-monotonic behavior. For $B<1.2$ T
a negative MR due to the formation of rosette trajectories is
observed. This NMR is then replaced by a non-saturating positive
MR, which is similar to the positive MR observed in the
unpatterned 2DEGs. A the same time the Hall resistance shows no
extra-features evolving from a curve with plateaus into strait
line when the temperature is increased (same behavior as in
unpatterned samples). Qualitatively the behavior is well described
by theory$^{9}$. However the detailed dependence of $\rho_{xx}(B)$
obtained from analytical calculations (see part III C from
Ref.~\onlinecite{Polyakov}) does not coincide with the
experimental data ; neither for the negative MR nor for the
positive MR and the theory fails to describe the details of our
data. We do not observe the $\rho_{xx}(B)\sim B^{-1}ln(B)$ for NMR
and the value $\alpha=12/7$ for PMR of the experimental curve near
$\rho_{xx}(B)$ minimum. The experiment shows $\rho_{xx}(B)\sim
B^{-1/2}$ and $\rho_{xx}(B)\sim B^{0.9}$. One of the possible
reasons of this disagreement is that the theory considers a random
array of antidots while our experiment is performed in a periodic
lattice.\\Anyway even if it fails to describe accurately our data
the theory predicts a minimum and it is interesting to compare the
position of the experimental minimum to this prediction. Indeed
its position depends on the correlation length of the long range
potential and of the average density of antidots independently of
the arrangement of the antidots. The theory gives the following
condition at the minimum:
\begin{center}
$\zeta/R_{cm}\approx p^{-4/19}[n\zeta^2ln(1/n\zeta^2p^{1/3})]^{7/19}$\\
\end{center}

Where $p=l_s/(l_L\zeta)^{1/2}$, $l_s$ is the mean free path in the
short range potential, $l_L$ is the mean free path in the long
range potential, $n$ is the density of antidots, $\zeta$ is the
correlation length of the long range potential and $R_{cm}$ is the
cyclotron radius at the minimum. For our lattice parameters we
obtain $\zeta\approx100 nm$. This value is close to the spacer
thickness ($60 nm$) which is known to determine the correlation
length of the long range potential.

\section{conclusion}
In conclusion , we have shown that the quasi-classical MR of a
high mobility 2DEG in AlGaAs/GaAs heterostructure is very large
and non-saturating up to values of  $\omega_c\tau$ exceeding 300.
The magnetic field dependence of  $\rho_{xx}(B)$ follows the law
$\rho_{xx}(B)\sim B^\alpha$ ($\alpha$=0.9-1.1). In antidot
lattices we observe a non-monotonic behavior of the MR (negative
first and then positive and non-saturating similar to that
observed in unpatterned samples). The comparison of the
experimental results with the theory$^{9}$ gives satisfactory
qualitative agreement but the theory fails to describe the details
of the magneto-resistance traces.

\begin{acknowledgments}
We thank A. Mirlin, D. Polyakov and I. Gornyi  for useful
discussions. This work has been supported by, the programs
PICS1577-RFBR02-02-22004, NATO Linkage (N CLG.978991), INTAS
(Grant N 01-0014) programs "Physics and Technology of
Nanostructures" of the Russian ministry of Industry and Science
and "Low dimensional quantum structures" of RAS.
\end{acknowledgments}

\end{document}